
\input phyzzx
\hoffset=0.2truein
\voffset=0.1truein
\hsize=6truein
\def\TITLEPAGE{\frontpagetrue}
\def\CALT#1{\hbox to \hsize{\tenpoint \baselineskip=12pt
	\hfil \vtop{
        \hbox{\strut hep-ph/9409319}
	\hbox{\strut CALT-68-#1}
	\hbox{\strut DOE RESEARCH AND}
	\hbox{\strut DEVELOPMENT REPORT}}}}
\def\CALTECH{
	\address{California Institute of Technology,
    Pasadena, CA 91125}}
\def\TITLE#1{\vskip.5in \centerline{\fourteenpoint#1}}
\def\AUTHOR#1{\vskip.2in \centerline{#1}}
\def\ABSTRACT#1{\vskip.2in \vfil \centerline
	    {\twelvepoint \bf Abstract}
		     #1 \vfil}
\def\ENDTITLEPAGE{\vfil \eject \pageno=1}
\hfuzz=5pt
\tolerance=10000
\TITLEPAGE
\CALT{1948}
\TITLE{Is Baryon Number Violated when Electroweak Strings Intercommute?}
\AUTHOR{Hoi-Kwong Lo\foot{Address after Sep. 1: School of
Natural Sciences, Institute for
Advanced Study, Princeton, NJ 08540}}
\CALTECH
\ABSTRACT{We reexamine the self-helicity and the intercommutation
of electroweak strings. A plausible argument for baryon number conservation
when electroweak strings intercommute is presented.
The connection between a segment of electroweak string and
a sphaleron is also discussed.}
\ENDTITLEPAGE
\eject
One of the most basic observational facts in cosmology is the predominance
of matter over antimatter. In 1967 Sakharov\Ref\Sak{A. D.
Sakharov, JETP Lett. {\bf 5}, 24 (1967).}
considered the possibility that the Universe began in a baryon-symmetric
state but that particle interactions produced a net asymmetry. In addition
to baryon-number violation, this requires $C$ and $CP$ violation as well
as a departure from thermal equilibrium.\Ref\Dim{S. Dimopoulos and
L. Susskind, Phys. Rev. {\bf D18}, 4500 (1978).}
In the standard electroweak theory all three conditions are satisfied.
Baryogenesis during the weak phase
transition\Ref\Coh{A. G. Cohen, D. B. Kaplan, and A. E. Nelson, Annu. Rev.
Nucl. Part. Sci. {\bf 43}, 27 (1993).}
is particularly interesting as it may eventually be experimentally
verifiable.

Recently, there has been a revival of interest in the study of classical
solutions in the standard model of the electroweak
interactions.\Ref\elec{T. Vachaspati and A. Ach\'ucarro, Phys. Rev.
{\bf D44}, 3067 (1991); T. Vachaspati, Phys. Rev. Lett. {\bf 68}, 1977 (1992).}
It has
been conjectured\REF\str{T. Vachaspati,
``Electroweak Strings: a Progress Report,``
in proceedings to Texas/PASCOS 92: Relativistic Astrophysics and Particle
Cosmology [Ann. N. Y. Acad. Sci. {\bf 688} (1993)], M. Barriola, T.
Vachaspati, and M. Bucher, Phys. Rev. D.
(to be published).}\REF\VacField{T. Vachaspati and G. B. Field, Phys. Rev.
Lett. {\bf 73}, 373 (1994).}\REF\Vachas{T. Vachaspati, Report No.
hep-ph/9405286, 1994
(to be published).}\refmark{\str,\VacField,\Vachas}
that a segment of electroweak string, which connects
a monopole to an antimonopole is a kind of ``stretched``
sphaleron.
Since
the sphaleron\REF\Man{N. S. Manton, Phys. Rev. {\bf D28}, 2019
(1983).}\REF\Klinkman{F. R. Klinkhamer and N. S. Manton, Phys. Rev. {\bf D30},
2212 (1984).}\refmark{\Man,\Klinkman}
is of crucial interest to the study of baryon number
violation processes, one may contemplate the role electroweak strings
in baryogenesis shortly after the cosmological electroweak phase
transition.\Ref\baryog{Baryogenesis scenarios based on
electroweak strings have been discussed by R. Brandenberger and A. C. Davis,
Phys. Lett. {\bf B308}, 79 (1993), M. Barriola, Report No.
hep-ph/9403323, 1994 (to be published), and
in Refs. 6 and 7. }

In a recent Letter,\refmark{\VacField} it is
pointed out that the baryon number anomaly equation may be interpreted
as a conservation law for baryon number minus helicity. Since the helicity
is a sum of link and twist numbers, linked and twisted loops of electroweak
string necessarily carry baryon number. It is also claimed
that helicity and hence baryon number is conserved when electroweak strings
intercommute.\Ref\bary{Baryon number violation may still occur
in other processes in the course of evolution of electroweak strings.}
However, two things put the argument given by those authors for
baryon number conservation during intercommutation in doubt.
First, the configuration after intercommutation given in Ref. 6
is implausible. Second, self-helicity is
not properly accounted for in that Letter.

The main point of this Letter is to give a careful calculation
of self-helicity and present a plausible electroweak
string configuration after intercommutation.
A proper calculation of helicity with our field
configuration after intercommutation suggests that helicity is conserved
when electroweak strings intercommute.
We also clarify the connection between the sphaleron and a segment of
$Z$ string connecting a monopole and an antimonopole.

We will briefly review the concept of helicity of electroweak
strings.\refmark{\VacField,\Vachas}
Our starting point is the Adler-Bell-Jackiw anomaly equation
$$
\partial_\mu J^\mu_B= {N_F \over 32 \pi^2} [g^2 W_a^{\mu \nu} \tilde W^a_{
\mu \nu} - g'^2 Y_{\mu \nu} \tilde Y^{\mu \nu} ], \eqno(1)$$
where $N_F$ is the number of families, $W_a^{\mu \nu}$ $a=1, 2,3$ and $Y_{
\mu \nu}$ are the $SU(2)$ and $U_Y(1)$ field strengths respectively. Note
that the right hand side can be expressed as a total divergence. We define
the Chern-Simons numbers $N_{CS}$ and $n_{CS}$
$$
\eqalignno{N_{CS}&\equiv {g^2 \over 32 \pi^2} \int d^3 x \epsilon^{ijk}
[W_{aij} W^a_k -{g \over3} \epsilon_{abc} W^{a}_i W^{b}_j W^{c}_k ]\cr
n_{CS}&\equiv {g'^2 \over 32 \pi^2} \int d^3 x \epsilon^{ijk} Y_{ij} Y_k,
&(2)\cr}$$
where $ W^a_k$ and $Y_k$ are gauge potentials
and obtain by integrating Eq. (1) over a spacetime volume that the
change in baryon number is related to the changes in the Chern-Simons numbers:
$$
\Delta  B=N_F (\Delta N_{CS} - \Delta n_{CS} ). \eqno(3)$$
The Chern-Simons number is not a meaningful physical
quantity as it changes by an integer upon a large gauge transformation.
However, the {\it change} in the Chern-Simons number in any physical process
is guage invariant. We will be interested in $Z$ string configurations
in which $W^1_\mu= W^2_\mu =0$.\Ref\gauge{Such a gauge choice is consistent
with the Lorentz gauge condition $\partial^\mu W^a_\mu =0$ and is valid for
{\it dynamical} $Z$ strings.}
With this simplification and the transformation
$$
\eqalignno{Z_j&= \cos \theta_w W^3_j - \sin \theta_w Y_j\cr
	  A_j&= \sin \theta_w W^3_j + \cos \theta_w Y_j . &(4)\cr}$$
Eq. (2) gives\Ref\Com{In Refs. 6 and 7, the factor 16 in the denominator
was mistaken to be 32.
We thank T. Vachaspati for pointing
this out. Because of another factor of 2 error of self-helicity to be
discussed below (See 17 below),
their calculation of the Chern-Simons number of
a twisted string remains valid.}
$$
N_{CS}-n_{CS}={\alpha^2 \over 16 \pi^2} \int d^3 x \bigl[
\cos (2 \theta_w) B_Z \cdot Z+ {1 \over 2 } \sin (2 \theta_w ) ( B_Z \cdot A
+ B_A \cdot Z) \bigr] \eqno(5)$$
where $\alpha= \sqrt{g^2+g'^2}$, $\tan \theta_w =g'/g$, $B$ denotes the
magnetic field, and the subscripts denote the gauge field for which
the magnetic field is to be evaluated.
As discussed by Vachaspati and Field, the first term on the r.h.s.
has a simple interpretation in terms of
the helicity\REF\hel{J. J. Moreau, C. R. Acad. Sci. Paris {\bf 252}, 2810
(1961); H. K. Moffatt, J. Fluid Mech. {\bf 35}, 117 (1969); M. Berger and
G. Field, J. Fluid Mech. {\bf 147}, 133
(1984).}\refmark{\hel,\VacField,\Vachas}
associated with the $Z$ field
$$H_Z= \int d^3 x B_Z \cdot Z . \eqno(6)$$
The helicity is a measure of the linkage of the magnetic field. If space
is divided into a collection of flux tubes, magnetic helicity arises from
the internal structure within each flux tube, such as twist and kinking, and
external relations between flux tubes, i.e. linking and
knotting.\Ref\source{With $W^1_\mu =W^2_\mu =0$, the $SU(2)$ gauge field
is Abelianized and the $Z$ flux does not terminate. Later in this paper
we will introduce monopoles and antimonopoles which rotate $Z$ flux to
$A$ flux. Configurations with monopoles and antimonopoles, however,
do not satisfy $W^1_\mu =W^2_\mu =0$.}
For two (untwisted) closed flux tubes linked once, a simple integration of
Eq. (6) gives
$$
H= \pm 2 \Phi_1 \Phi_2 \eqno(7)$$
where $\Phi_1$ and $\Phi_2$ measure the magnetic fluxes of the tubes and
the sign of $H$ depends on the sense of linkage. This ends our review
of the discussion of helicity made by Vachaspati and Field.

Consider a $Z$ string loop of flux $\Phi$ that is twisted by an angle
$2p \pi$.
To compute its (internal) helicity,\Ref\RicMof{See e.g. R. I. Ricca and H.
K. Moffatt, in {\it Topological Aspects of the Dynamics of Fluids and Plasma},
edited by H. K. Moffatt {\it et al}, NATO ASI series E Vol. 218
(Kluwer Academic Publishers, 1992).} let us divide the tube up into $m$
``subtubes`` each with the flux $\Phi/m$. The linking number of each pair
of subtubes is the same as that of each pair of magnetic field lines in the
loop. We observe that, for a pair of field lines in a uniformly twisted
torus, one of the field lines can always be deformed to the axis of the tube
without intersecting the second field line. Thus, the linking number is just
$p$. The total helicity is the sum of these self-helicities plus the
interactive helicities arising from the linkage of the flux tubes. Therefore,
$$H=m H_m +2 \sum_{i <j} p \Phi_i \Phi_j , \eqno(8)$$
with $\Phi_i= \Phi/m$ ($i=1,2,\dots,m$). Since self-helicities scale as
$\Phi^2$,
$$H={H \over m} +2p {m(m-1) \over 2} \left( {\Phi \over m} \right)^2
\eqno(9)$$
i.e.\Ref\error{Refs. 6 and 7 erroneously gave
$2p \Phi^2$ as the answer.} $$H=p\Phi^2. \eqno(10)$$

Let us turn to the main subject of our investigation: does intercommutation
of electroweak strings violate helicity (and hence baryon number)?
It is well known to fluid dynamicists that
helicity is approximately conserved in the
magnetohydrodynamics of
fluids with negligible viscosities and large magnetic Reynolds
numbers.\Ref\ren{The magnetic Reynolds number $R_m$ is defined to be
$ \mu_0 \sigma u_0 l_0$
where $\sigma$ is the electric conductivity of the fluid, $u_0$ is a
typical scale for the velocity field and $l_0$ is a typical length scale
over which it varies.} However, a network of electroweak string is far from
a superconducting fluid. If viscous effects are negligible, a fluid
can support no tangential stress. On the other hand, a string has
string tension.

Generally speaking, naive reconnections of
flux tubes after intercommutation
violate
helicity.
Conservation of helicity would require (additional local) twisting
(and/or writhing)
of flux tubes after reconnections.\Ref\AkhRuz{A.
Ruzmaikin and P. Akhmetiev, in {\it
Topological Aspects of the Dynamics of Fluids and Plasma},
edited by H. K. Moffatt {\it et al}, NATO ASI series E Vol. 218
(Kluwer Academic Publishers, 1992); Phys. Plasmas {\bf 1}, 331 (1994).}

The question is: what happens in general when two electroweak strings
intersect and intercommute? Consider a pair of antiparallel electroweak
strings approaching each other. The magnetic field goes through zero.
It breaks and reconnects immediately. Such a matching of
field lines leads to
helicity conservation during intercommutation.\Ref\bordism{The mathematical
framework of ``framed bordisms'' is introduced in Ref. 19 in the discussion of
the
reconnection process. Note that, owing to their erroneous calculation of
self-helicities,
the final field configuration suggested by
Vachaspati and Field actually violates helicity and is different from ours.}.
Pfister and Gekelman have proposed the following visual demonstration of
helicity conservation with a simple Christmas
ribbon.\Ref\Christmas{H. Pfister and W.
Gekelman, Am. J. Phys. {\bf 59}, 497 (1991).}
Let us begin with two singly linked loops of untwisted ribbons as shown in
Fig. 1(a).
We arrange
both ribbons such that they touch at one point and the $Z$ fields are
antiparallel there (as required for reconnection). We staple the ribbons
together to the left and right of the contact point and cut the ribbons
in between the staples (Fig. 1(b)).
What we have obtained is one loop that has
two complete ($360^\circ$) twists as shown in Fig. 1(c). From equation
(7) and (10), we find that both the initial and final configurations
have a helicity of $2 \Phi^2$. Therefore, helicity is conserved.
For comparison, we also show
the intercommutation of two unlinked string loops in Fig. 2.
They intercommute
to form an untwisted loop of string. Even though we have only discussed
the case of two antiparallel electroweak strings, we conjecture
that helicity conservation is valid for electroweak strings intersecting
at any angle. The magnetic field always goes through zero and
reconnect immediately as in the case of perfect MHD.

A segment of $Z$ string can also exist, but it has
to connect a monopole ($m$) to
an antimonopole ($\bar m$).
The asymptotic
field configurations of $m$ and $\bar m$ (each connected to a semi-infinite
$Z$ string)
have been written by
Nambu\Ref\Nambu{Y. Nambu, Nucl. Phys. {\bf B130}, 505 (1977). $\Phi$ has
been rescaled so that the vacuum manifold is given by $\Phi^{\dag} \Phi=1$.}
$$
\Phi_m= \left( {\cos (\theta_m /2) \atop \sin (\theta_m /2) e^{i \phi}}
\right),\qquad \Phi_{\bar m}= \left( {\sin (\theta_{\bar m} /2) \atop
\cos (\theta_{\bar m} /2 ) e^{i \phi} } \right), \eqno(11)$$
where $\theta_m$ and $\phi$ are spherical
coordinates centered on $m$ (and similarly for $\bar m$) and
the gauge field satisfies
$D_i \Phi =0$.\Ref\zflux{This involves a gauge choice and assumes that
there is no $A$ flux in the semi-infinite flux tube.}
It has been suggested\refmark{\str}
that a segment of $Z$ string
is a kind of
``extended'' sphaleron.
Here we are interested in a ``family'' of extended sphaleron
by which we mean a set of field
configurations each with a Chern-Simons number of $1/2$
parametrized by a deformation variable, $d$, such that the
$d=0$ element is a sphaleron.
However, as discussed by M. Hindmarsh,\Ref\Hind{M.
Hindmarsh, Report No. hep-ph/9408241.} a simple symmetry argument shows
that an untwisted segment of $Z$ string has a Chern-Simons number zero
rather than $1/2$.
More
recently, Vachaspati and Field\refmark{\VacField}
have considered the Higgs field configuration
$$
\Phi_{m \bar m} (\gamma)= \left( {\sin (\theta_m /2) \sin (\theta_{\bar m}/2)
e^{i \gamma} + \cos ( \theta_m/2) \cos ( \theta_{\bar m}/2) \atop
\sin (\theta_m /2) \cos ( \theta_{\bar m} /2 ) e^{i \phi} - \cos (\theta_m /2)
\sin (\theta_{\bar m}/2 ) e^{i (\phi - \gamma)}} \right), \eqno(12)$$
where $\theta_m$ and $\theta_{\bar m}$ are the polar angles and $\phi$ is
the aximuth angle.
Eq. (12) reduces to $\Phi_m$ when $\theta_{\bar m} \to 0$
and to $e^{i \gamma} \Phi_{\bar m}$ when $\theta_m \to \pi$ and, in addition,
we perform the rotation $\phi \to \phi + \gamma$. Thus, we see that the
antimonopole is rotated and globally transformed by $U(1)$.
When $\theta_m \to \pi$ and $\theta_{\bar m} \to 0 $, Eq. (12) reduces
to the configuration of an {\it untwisted} $Z$ string.
Therefore, Eq. (12) describes an untwisted
segment of
$Z$ string connected to a monopole and a transformed antimonopole.
See Fig. 1e of Ref. 6.

In Refs. 6 and 7
this configuration was also loosely interpreted
as ``a monopole and antimonopole connected by a $Z$
string that is twisted through an angle $\gamma$''. We disagree with
such an interpretation. Their reinterpreted configuration
is, in fact, Fig. 1d of Ref. 6.
Starting with the field configuration of Fig. 1e, we can obtain
the configuration of Fig. 1d by the following process. Divide
the space into three regions, $x <-a$, $ -a <x <a$ and $x>a$.
In the first region, the two field configurations are the same.
In the third region, Fig. 1d is obtained by rotating Fig. 1e by $\gamma$.
(We are considering the general case where the twist is $\gamma$.)
In the second region, the string is twisted. In other words,
at each $x$, the configuration of Fig. 1d is locally
the result of rotating Fig. 1e by an angle , $\psi$,
a function of
$x$ such that $\psi (x=-a) =0$ and $\psi (x=a)= \gamma$.
It is easy to see that the contributions from regions 1 and 3
to the Chern-Simons
numbers of Fig. 1d and 1e
are the same. In the second region, the untwisted string of
Fig. 1e is replaced by one with a twist by $\gamma$. To
compute the change of Chern-Simons number, we must not forget
that since $B_A$ is non-zero in the presence of
monopoles, there are
contributions from the last two terms on the right hand
side of Eq. (5).\Ref\thanks{We are grateful to T. Vachaspati for
pointing this out.}
These two terms can be interpreted as the
cross linkage of $B_A$ with $B_Z$. Ignoring these terms
for a moment, from Eq. (6) and
(10) (with $\Phi= 4 \pi / \alpha$), the first term of Eq. (5) alone
will give a difference of the Chern-Simons numbers
${\alpha^2 \over 16 \pi^2} \cos (2 \theta_w){\gamma \over 2 \pi}
\left( {4 \pi \over \alpha} \right)^2 = {\gamma \over 2 \pi}
\cos (2 \theta_w) .$ Let us now return to the cross linkage piece.
Conservation of the hypercharge flux implies that\Ref\dyon{T. Vachaspati,
Report No. hep-ph/9405285, 1994 (to be published).}
the $B_A$ flux flowing across a plane $x= cst$ (for $-a <x <a$) is
${4 \pi \over \alpha} \tan \theta_w$. Hence, from Eqs. (5), (6)
and (7), the contribution to the change of C-S number from
cross linkage is $2 {\alpha^2 \over 16 \pi^2} ({1 \over2}) \sin (2 \theta_w)
{\gamma \over 2 \pi}
2({4 \pi \over \alpha})({4 \pi \over \alpha}) \tan \theta_w= {2 \gamma
\over \pi} \sin^2 \theta_w .$ Summing the two contributions gives
${\gamma \over 2 \pi} (1+ 2 \sin^2 \theta_w)$.
Note that when the monopole and antimonopole at the two ends of a
twisted segment of
$Z$ string are brought together
and annihilate each other, a closed
loop is formed and we expect $B_A$ to go away.

The authors of Refs. 6 and 7 attempted to compute the Chern-Simons number
of Fig. 1e. Unfortunately, they wrongly believed that the
Chern-Simons number of Fig. 1e is the same as that of Fig. 1d.
So, they argued as follows.
If $\gamma=2 \pi n /m$, where $n$ and $m$ are integers, one can join
together $m$ of these twisted segments and form a loop of $Z$ string that is
twisted by an angle $2 \pi n$.
They concluded
that the Chern-Simon number of the configuration
in Eq. (12) is ${\gamma \over 2 \pi}
 \cos (2 \theta_w)$ and for
$\gamma= \pi / {\cos (2 \theta_w)}$ it has a
Chern-Simons number of $1/2$. They conjectured that it is a ``stretched''
sphaleron. This conjecture is not well motivated because as discussed
in the last paragraph, the two configurations shown in Figs. 1d and 1e have
different Chern-Simons numbers. What
they have computed is actually the {\it contribution of the first term} of
Eq. (5)
to the {\it difference} of
the Chern-Simons numbers
of these two configurations
rather than that of Fig. 1e.

One can obtain an extended sphaleron
by setting $\gamma = \pi$ and considering parity odd field configurations.
In fact, there is a general
theorem by
Axenides and Johansen\Ref\AxeJoh{M. Axenides and A. Johansen, Mod. Phys.
Lett. {\bf A9}, 1033 (1994).} which states
that any parity odd configuration which
as $r \to \infty$ approaches the field configuration of the
sphaleron (at $\theta_w =0$)
have
($1/2$) as its Chern-Simons number.

The connection between a segment of electroweak string and a sphaleron
is considered. Besides,
we have discussed a plausible final field configuration of electroweak
strings when they intersect and intercommute and presented a careful
calculation of self-helicities. Taken together, they suggest that
baryon number is conserved when electroweak strings intercommute.
However, other processes in the course
of evolution of electroweak strings may still lead to
baryon number violation. An estimate for helicity density fluctuations
produced during the electroweak phase transition has been made in Ref. 6.
Suppose that a network of $Z$ string was produced at the electroweak
phase transition and it survives long enough to fall out of
thermal equilibrium. Since the full electroweak Lagrangian, which governs
the evolution of the system, is $CP$ violating, a change of helicity
in one direction is favored over the other and hence net change in
baryon number may occur.
\bigskip

We are indebted to T. Vachaspati for many stimulating
discussions and to J. Preskill for a critical reading of the manuscript.
We also acknowledge helpful discussions with
U. Aglietti, M. Axenides, M. Bucher, A. Y. K. Chui,
M. James and C. L. Y.
Lee. This work was
supported in part by the U. S. Department of Energy under Grant No.
DE-FG03-92-ER40701.
\refout
\bigskip
{\bf Figure Captions}:

FIG. 1. In the reconnection of two singly linked Christmas ribbons,
local twists are formed and helicity is conserved.

FIG. 2. Two unlinked Christmas ribbons intercommute to form an untwisted
ribbon.
\bye